\documentstyle[12pt,twoside]{article}
\begin{document}

\newcommand{\beqn}{\begin{eqnarray}}
\newcommand{\eeqn}{\end{eqnarray}}
\newcommand{\be}{\begin{equation}}
\newcommand{\ee}{\end{equation}}
\newcommand{\ba}{\begin{array}}
\newcommand{\ea}{\end{array}}
\newcommand{\ov}{\overline}
\newcommand{\ve}{\varepsilon}
\newcommand{\ds}{\displaystyle}

\newcommand {\N}{{\rm\bf N}}
\newcommand {\R}{{\rm\bf R}}
\newcommand{\T}{{\rm\bf T}}
\newcommand {\Z}{{\rm\bf Z}}
\newcommand {\C}{{\rm\bf C}}

\renewcommand{\theequation}{\thesection.\arabic{equation}}

\newcommand{\const}{\mathop{\rm const}\nolimits}
\newcommand{\dist}{\mathop{\rm dist}\nolimits}
\newcommand{\tr}{\mathop{\rm tr}\nolimits}
\newcommand{\supp}{\mathop{\rm supp}\nolimits}
\newcommand{\sign}{\mathop{\rm sign}\nolimits}
\newtheorem{theorem}{Theorem}[section]
\renewcommand{\thetheorem}{\arabic{section}.\arabic{theorem}}
\newtheorem{definition}[theorem]{Definition}
\newtheorem{lemma}[theorem]{Lemma}
\newtheorem{example}[theorem]{Example}
\newtheorem{remark}[theorem]{Remark}
\newtheorem{remarks}[theorem]{Remarks}
\newtheorem{cor}[theorem]{Corollary}
\newtheorem{pro}[theorem]{Proposition}

\newcommand{\bo}{{\hfill\loota}}
\newcommand{\loota}{\hbox{\enspace{\vrule height 7pt depth 0pt width 7pt}}}

\begin{titlepage}
\begin{center}
{\Large\bf 
Harmonic Crystals in the Half-Space, I.\\
\vspace{3mm}
Convergence to Equilibrium}\\
\vspace{2cm}
{\large T.V.~Dudnikova}
\footnote{Supported partly by
research grant of RFBR (06-01-00096)}\\
{\it Elektrostal Polytechnical Institute\\
 Elektrostal 144000, Russia}\\ 
e-mail:~dudnik@elsite.ru
\end{center}
 \vspace{1cm}
 \begin{abstract}
We consider the dynamics of a harmonic crystal in the half-space
with zero boundary condition. It is assumed that the initial
date is a  random function  with zero mean, finite mean energy density 
which also satisfies  a  mixing condition of Rosenblatt or Ibragimov type.
We study the distribution $\mu_t$  of the  solution at time $t\in\R$.
The main result is the convergence of $\mu_t$ to a Gaussian measure 
as $t\to\infty$ which is time stationary with a covariance 
inherited from the initial (in general, non-Gaussian) measure. 
 \end{abstract}

\end{titlepage}
 \section{Introduction}

The paper concerns the problems of the long-time convergence
to the equilibrium distribution for the discrete systems.
For one-dimensional chains of harmonic oscillators the problem is analyzed 
in \cite{BPT, SL}: in  \cite{SL} -- for initial measures
which have distinct temperatures to the left and to the right, 
and in \cite{BPT} -- for a more general class of initial measures 
characterized by a mixing condition of  Rosenblatt- or Ibragimov- type 
and which are asymptotically translation-invariant to the left 
and to the right.
For many-dimensional harmonic crystals the  convergence has been proved
in \cite{LL} for initial measures which are absolutely continuous with 
respect to the canonical Gaussian measure. In  \cite{DKKS}--\cite{DK2}
we have started the convergence analysis for partial 
differential equations of hyperbolic type in $\R^d$, $d\ge1$.
In \cite{DKS1}--\cite{DS}  we extended the results to harmonic crystals. 
In the harmonic approximation the crystal is characterized by the
displacements $u(z,t)\in\R^n$,  $z\in\Z^d$, of the crystal atoms from their
equilibrium positions. The field  $u(z,t)$ is governed by a discrete 
wave equation.
In the papers  mentioned above the lattice dynamics has been studied in
the whole space $\Z^d$.
 
In the present work the dynamics of the harmonic crystals is studied
 in the half-space  $\Z^d_+$, $d\ge 1$, 
\be\label{1+}
\ddot u(z,t)=-\sum\limits_{z'\in \Z^d_+}\left(V(z-z')-V(z-\tilde z')\right)
u(z',t),\,\,\,\,z\in\Z^d_+,\,\,\,\,t\in\R,
\ee
where $\tilde z:=(-z_1,\bar z)$, $\bar z=(z_2,\dots,z_d)\in \Z^{d-1}$,
with zero boundary condition,
\be\label{2+}
u(z,t)|_{z_1=0}=0,
\ee
and with the initial data
\be\label{3+}
u(z,0)=u_0(z),\quad \dot u(z,0)=u_1(z),\quad z\in\Z^d_+.
\ee
Here $\Z^d_+=\{z\in \Z^d:\,z_1>0\}$,
$V(z)$ is the interaction (or force) matrix, $\left(V_{kl}(z)\right)$,
$k,l=1,\dots,n$,  $u(z,t)=(u_1(z,t),\dots,u_n(z,t))$,
$u_0(z)=(u_{01}(z),\dots,u_{0n}(z))\in\R^n$ and correspondingly for $u_1(z)$. 
To coordinate the boundary and initial conditions we suppose
that  $u_0(z)=u_1(z)=0$ for $z_1=0$.

Denote $Y(t)=(Y^0(t),Y^1(t))\equiv (u(\cdot,t),\dot u(\cdot,t))$,
$Y_0=(Y_0^0,Y_0^1)\equiv (u_0(\cdot),u_1(\cdot))$. Then
(\ref{1+})--(\ref{3+}) takes the form of the evolution equation
\be\label{CP1}
\dot Y(t)={\cal A}_+Y(t),\quad t\in\R,\,\,z\in\Z^d_+,
\quad Y^0(t)|_{z_1=0}=0,\quad Y(0)=Y_0.
\ee
Here ${\cal A}_+=\left(\ba{cc}0&1\\-{\cal V}_+&0\ea\right)$,
with 
${\cal V}_+u(z):= \sum\limits_{z'\in\Z^d_+}(V(z-z')-V(z-\tilde z'))u(z')$.

It is  assumed that the initial state $Y_0$ is given by a
random element of the Hilbert space ${\cal H}_{\alpha,+}$ of real sequences,
see Definition \ref{d1.1} below.
  The distribution of $Y_0$ is a probability measure  $\mu_0$
 satisfying conditions {\bf S1}--{\bf S4} below.
In particular, the initial correlation function 
$Q_0(z,z')$ is asymptotically translation-invariant
 as $z_1,z'_1\to+\infty$ (see Condition {\bf S2}) and the measure $\mu_0$
has some mixing properties (see Condition {\bf S4}).  
Given $t\in\R$, denote by $\mu_t$ the probability measure
on ${\cal H}_{\alpha,+}$ giving the distribution of the random 
solution $Y(t)$ to the problem (\ref{CP1}).

Our main result gives the weak convergence of measures
$\mu_t$ on the space ${\cal H}_{\alpha,+}$, with $\alpha<-d/2$,
 to a limit measure $\mu_{\infty}$,
\be\label{1.8i}
\mu_t \,\buildrel {\hspace{2mm}{\cal H}_{\alpha,+}}\over
{- \hspace{-2mm} \rightharpoondown }
\mu_\infty\quad{\rm as}\,\,\,\, t\to \infty,
\ee
where $\mu_\infty$ is an equilibrium Gaussian measure on 
${\cal H}_{\alpha,+}$.  This means the convergence
$$
 \int f(Y)\mu_t(dY)\rightarrow
 \int f(Y)\mu_\infty(dY),\quad t\to \infty,
 $$
 for any bounded continuous functional $f$
 on ${\cal H}_{\alpha,+}$.

Explicit formulas for the correlation functions of the measure $\mu_0$
are given in (\ref{1.13})--(\ref{1.15}).

The paper is organized as follows.
The conditions on the interaction matrix $V$ and the initial measure $\mu_0$
are given in Section~2. The main result is stated in Section 3.
Examples of harmonic crystals and the initial measures satisfying 
all conditions imposed are constructed in Section 4.
The convergence of correlation functions of $\mu_t$ 
is established in Section 5, the compactness of $\mu_t$, $t\ge0$,
and the  convergence of characteristic functionals of $\mu_t$
are proved in Sections 6 and 7, respectively.

\setcounter{equation}{0}
 \section{Conditions on the system and the initial measure}
 \subsection{Dynamics}

Let us assume that
\be\label{condE0}
V(z)=V(\tilde z),\quad \mbox{where }\,\tilde z:=(-z_1,\bar z),
\quad\bar z=(z_2,\dots,z_d)\in \Z^{d-1}.
\ee
Then the solution to the problem  (\ref{1+})--(\ref{3+}) 
can be represented as the restriction of the solution to the Cauchy problem
with the odd initial date on the half-space. 
More precisely, consider the following Cauchy problem 
for the harmonic crystal in the whole space $\Z^d$:
\beqn\label{CP1'}
\left\{\ba{l}
\ddot v(z,t)=-\sum\limits_{z'\in \Z^d}V(z-z')v(z',t),\,\,\,\,z\in\Z^d,
\,\,\,\,t\in\R,\\
v(z,0)=v_0(z),\quad \dot v(z,0)=v_1(z),\quad z\in\Z^d.
\ea\right.
\eeqn
Denote $X(t)=(X^0(t),X^1(t))\equiv (v(\cdot,t),\dot v(\cdot,t))$,
$X_0=(X_0^0,X_0^1)\equiv (v_0(\cdot),v_1(\cdot))$. Then
(\ref{CP1'}) has a form 
\be\label{CP1''}
\dot X(t)={\cal A}X(t),\quad t\in\R,\quad X(0)=X_0.
\ee
Here ${\cal A}=\left(\ba{cc}0&1\\-{\cal V}&0\ea\right)$,
where ${\cal V}$ is a convolution operator with the matrix kernel $V$.

 Let us assume that the initial date $X_0(z)$ 
be an odd function w.r.t. $z_1\in\Z^1$, i.e.,
$X_0(z)=-X_0(\tilde z)$.
Then the solution $v(z,t)$ of (\ref{CP1'})
is also an odd function w.r.t. $z_1\in\Z^1$.
Let us restrict the solution $v(z,t)$ 
 on the domain $\Z^d_+$ and put $u(z,t)=v(z,t)|_{z_1\ge0}$.
Then $u(z,t)$ is the solution to the problem (\ref{CP1}) with
the initial date $Y_0(z)=X_0(z)|_{z_1\ge0}$. 

  Assume that the initial date $Y_0$ of the problem (\ref{CP1})
belongs to the phase space ${\cal H}_{\alpha,+}$, $\alpha\in\R$,
defined below.
 \begin{definition}                 \label{d1.1}
 $ {\cal H}_{\alpha,+}$ is the Hilbert space
of $\R^n\times\R^n$-valued functions  of $z\in\Z^d_+$
 endowed  with  the norm
 \beqn\nonumber 
 \Vert Y\Vert^2_{\alpha,+}
 = \sum_{z\in\Z^d_+}\vert Y(z)\vert^2(1+|z|^2)^{\alpha} <\infty.
 \eeqn  
\end{definition}
In addition it is assumed that the initial date $Y_0=0$ if $z_1=0$.

We impose the following conditions {\bf E1}--{\bf E6} on the matrix $V$.
\medskip\\
{\bf E1}. There exist positive constants $C,\gamma$ such that
$\|V(z)\|\le C e^{-\gamma|z|}$ for $z\in \Z^d$,
$\|V(z)\|$ denoting the matrix norm.
\medskip

Let $\hat V(\theta)$ be the Fourier transform of $V(z)$, with the
convention
 $$
 \hat V(\theta)=
\sum\limits_{z\in\Z^d}V(z)e^{iz\cdot\theta}\,,\theta \in \T^d,
 $$
where "$\cdot$" stands for the scalar product in Euclidean space $\R^d$
and  $\T^d$ denotes the $d$-torus $\R^d/(2\pi \Z)^d$.
\medskip\\
{\bf E2}. $ V$ is real and symmetric, i.e., $V_{lk}(-z)=V_{kl}(z)\in \R$,
$k,l=1,\dots,n$, $z\in \Z^d$.
\medskip\\
Both conditions imply that $\hat V(\theta)$ is a real-analytic
 Hermitian matrix-valued function in $\theta\in \T^d\!$.
\medskip\\
{\bf E3}. The matrix $\hat V(\theta)$ is  non-negative definite for
every $\theta \in \T^d$.
\medskip

Let us define the Hermitian  non-negative definite matrix,
 \be\label{Omega}
 \Omega(\theta)=\big(\hat V(\theta )\big)^{1/2}\ge 0.
 \ee
$\Omega(\theta)$  has the eigenvalues 
$0\leq\omega_1(\theta)<\omega_2(\theta) \ldots <\omega_s(\theta)$, 
$s\leq n$, and the
corresponding spectral projections $\Pi_\sigma(\theta)$ with
multiplicity $r_\sigma=\tr\Pi_\sigma(\theta)$. $\theta
\mapsto\omega_\sigma(\theta)$ is the $\sigma\!$-th band function.
There are special points in $\T^d$,
where the bands cross, which means that $s$ and $r_\sigma$ jump to
some other value. Away from such crossing points $s$ and
$r_\sigma$ are independent of $\theta$. More precisely one has the
following lemma.
\begin{lemma}\label{lc*} (see \cite[Lemma 2.2]{DKS1}).
Let the conditions {\bf E1} and {\bf E2}
hold. Then there exists a closed subset ${\cal C}_*\subset \T^d$
such that we have the following:\\
(i) the Lebesgue measure of ${\cal C}_*$ is zero.\\
(ii) For any point $\Theta\in \T^d\setminus{\cal C}_*$ there
exists a neighborhood ${\cal O}(\Theta)$ such that each band
function $\omega_\sigma(\theta)$ can be chosen as the real-analytic function in
${\cal O}(\Theta)$.\\
(iii) The eigenvalue $\omega_\sigma(\theta)$ has constant multiplicity
in $\T^d\setminus{\cal C}_*$.\\
(iv) The spectral decomposition holds, 
 \be\label{spd'}
\Omega(\theta)=\sum_{\sigma=1}^s \omega_\sigma
(\theta)\Pi_\sigma(\theta),\quad \theta\in \T^d\setminus{\cal C}_*, 
 \ee
 where $\Pi_\sigma(\theta)$ is the orthogonal projection in
$\R^n$. $\Pi_\sigma$ is a real-analytic function on
$\T^d\setminus{\cal C}_*$.
\end{lemma}

For $\theta\in \T^d\setminus{\cal C}_*$, we denote by
Hess$(\omega_\sigma)$ the matrix of second partial derivatives. 
The next condition on $V$ is the following:
\smallskip\\
{\bf E4}. Let $D_\sigma(\theta)=\det\big(\rm{Hess}(\omega_\sigma(\theta))\big)$.
Then $D_\sigma(\theta)$ does not vanish identically on
$\T^d\setminus{\cal C}_*$, $\sigma=1,\ldots,s$.
\medskip

Let us denote
 \be\label{c0ck}
{\cal C}_0=\{\theta\in \T^d:\det \hat
 V(\theta)=0\}\,\, \mbox{and }\,
{\cal C}_\sigma=\{\theta\in \T^d\setminus {\cal
C}_*:\,D_\sigma(\theta)=0\},\,\,\, \sigma=1,\dots,s.
 \ee
Then the Lebesgue measure of ${\cal C}_\sigma$ vanishes, $\sigma=0,1,...,s$
(see \cite[Lemma 2.3]{DKS1}).

The last conditions on $V$ are the following:
\medskip\\
{\bf E5}.  For each $\sigma\ne \sigma'$, 
the identities $\omega_\sigma(\theta)
\pm\omega_{\sigma'}(\theta)\equiv\const_\pm$, 
$\theta\in \T^d\setminus {\cal C}_*$, do not hold  with $\const_\pm\ne 0$.
\medskip\\
This condition holds trivially  in the  case $n=1$.
\medskip\\
{\bf E6}. $\Vert \hat V^{-1}(\theta)\Vert\in L^1(\T^d)$.\medskip\\
If ${\cal C}_0=\emptyset$, then $\|\hat{V}^{-1}(\theta)\|$ is
bounded and {\bf E6} holds trivially.
\medskip

Denote by $ {\cal H}_\alpha$ the Hilbert space
of $\R^n\times \R^n$-valued functions  of $z\in\Z^d$
 endowed  with  the norm
$$
 \Vert X\Vert^2_{\alpha} = \sum_{z\in\Z^d}\vert X(z)\vert^2
(1+|z|^2)^{\alpha} <\infty.
$$
\begin{pro} \label{p1.1} (see \cite[Proposition 2.5]{DKS1}).
Let conditions {\bf E1} and {\bf E2} hold, and choose some $\alpha\in\R$. 
Then (i) for any  $X_0 \in {\cal H}_\alpha$,
 there exists  a unique solution $X(t)\in C(\R, {\cal H}_\alpha)$
 to the Cauchy problem (\ref{CP1''}).\\
(ii) The operator $U(t):X_0\mapsto X(t)$ is continuous
in ${\cal H}_\alpha$.
\end{pro}
\begin{cor}\label{c1}
Let conditions (\ref{condE0}), {\bf E1} and {\bf E2} hold. Then
(i) for any  $Y_0 \in {\cal H}_{\alpha,+}$, there exists  a unique solution
$Y(t)\in C(\R, {\cal H}_{\alpha,+})$  to the mixed problem (\ref{CP1}).\\
(ii) The operator  $U_+(t):Y_0\mapsto Y(t)$ is continuous
in ${\cal H}_{\alpha,+}$.
\end{cor}
{\bf Proof}. 
Corollary \ref{c1} follows from Proposition \ref{p1.1}. Indeed,
 the solution  $X(z,t)$ of (\ref{CP1''}) admits the representation   
\be\label{solGr}
X(z,t)=\sum\limits_{z'\in\Z^d}{\cal G}_t(z-z')X_0(z'),
\ee
where the Green function ${\cal G}_t(z)$ has the Fourier representation   
\be\label{Grcs}  
{\cal G}_t(z):= F^{-1}_{\theta\to z}[  
\exp\big(\hat{\cal A}(\theta)t\big)]  
=(2\pi)^{-d}\int\limits_{\T^d}e^{-iz\cdot\theta}  
\exp\big(\hat{\cal A}(\theta)t\big)\,d\theta  
\ee  
with 
\be\label{hA}   
\hat{\cal A}(\theta)=\left( \begin{array}{cc}   
0 & 1\\   
-\hat V(\theta) & 0   
\end{array}\right),\,\,\,\,\theta\in \T^d.    
\ee   
Therefore, the solution to the problem (\ref{CP1}) has a form
\be\label{sol}
Y(z,t)=\sum\limits_{z'\in\Z^d_+} {\cal G}_{t,+}(z,z')
Y_0(z'),\quad z\in\Z^d_+,
\ee
where ${\cal G}_{t,+}(z,z'):=
{\cal G}_t(z-z')-{\cal G}_t(z-\tilde z')$.
Corollary \ref{c1} follows. \bo

\subsection{Random initial data and statistical conditions}

Denote by $\mu_0$ a Borel probability measure   
on ${\cal H}_{\alpha,+}$ giving the distribution of $Y_0$.
Expectation with respect to $\mu_0$  is denoted by $E$.

Assume that the initial measure $\mu_0$  
has the following properties.
\medskip\\
{\bf S1}. $Y_0(z)$ has zero expectation value,
$E_0 \big(Y_0(z)\big) = 0$, $z\in\Z^d_+$.

For $a,b,c \in \C^n$,  denote by $a\otimes b$ the
 linear operator $(a\otimes b)c=a\sum^n_{j=1}b_j c_j$.\\
{\bf S2}. The correlation matrices of the measure $\mu_0$ have a form
 \beqn \label{1.9'}
Q^{ij}_0(z,z')= E_0\big(Y^i_0(z)\otimes {Y^j_0(z')}
\big)=  q^{ij}_0(z_1,z'_1,\bar z-\bar z'),\,\,\,z,z'\in\Z^d_+,\,\,\,i,j=0,1.
 \eeqn
where \\
(i) $q^{ij}_0(z_1,z'_1,\bar z)=0$ for $z_1=0$ or $z_1'=0$,\\
(ii) $\lim_{y\to+\infty} q^{ij}_0(z_1+y,y,\bar z)
={\bf q}_0^{ij}(z)$, $z=(z_1,\bar z)\in\Z^d$.
Here ${\bf q}_0^{ij}(z)$ are correlation functions of some 
translation invariant measure $\nu_0$ with zero mean value 
in ${\cal H}_{\alpha}$.
\begin{definition}
A measure $\nu$ is called translation invariant if 
$\nu(T_h B)= \nu(B)$,  $B\in{\cal B}({\cal H}_{\alpha})$,
$h\in\Z^d$, where $T_h X(z)= X(z-h)$, $z\in\Z^d$.   
\end{definition}
{\bf S3}. The measure $\mu_0$  has a finite variance 
and finite mean energy density,
 \beqn\label{med}
e_0(z)=E_0 \big(\vert Y_0^0(z)\vert^2
 + \vert Y_0^1(z)\vert^2\big)
=\tr\left[Q_0^{00}(z,z)+Q_0^{11}(z,z)\right]\le e_0<\infty,\,\,\,z\in\Z^d_+.
 \eeqn

Finally, it is assumed that the measure $\mu_0$ satisfies a mixing condition.
To formulate this condition, let us denote by $\sigma ({\cal A})$,
${\cal A}\subset \Z^d_+$, the $\sigma $-algebra in
${\cal H}_{\alpha,+}$ generated by $Y_0(z)$ with $z\in{\cal A}$.
Define the Ibragimov mixing coefficient of the probability  measure
$\mu_0$ on ${\cal H}_{\alpha,+}$ by the rule (cf. \cite[Definition 17.2.2]{IL})
 \beqn \label{ilc} 
\varphi(r)= \sup_{\scriptsize{\ba{cc} {\cal A},{\cal B}\subset \Z^d_+\\
\dist({\cal A},\,{\cal B})\geq r \ea}}
\sup_{\scriptsize{
\ba{cc} A\in\sigma({\cal A}),B\in\sigma({\cal B})\\ \mu_0(B)>0\ea}}
\frac{|\mu_0(A\cap B) - \mu_0(A)\mu_0(B)|}{ \mu_0(B)}.
 \eeqn
\begin{definition}
 A measure $\mu_0$ satisfies the strong uniform Ibragimov mixing condition if
$\varphi(r)\to 0$ as $r\to\infty$.
\end{definition}
{\bf S4}. The measure $\mu_0$ satisfies the strong uniform
Ibragimov mixing condition with
 \be\label{1.12}
\int\limits_{0}^{\infty}
 r^{d-1}\varphi^{1/2}(r)\,dr <\infty\,.
 \ee
This condition can be considerably weakened 
(see Remarks \ref{remi-iii} (i), (ii)).

\setcounter{equation}{0}
 \section{Main results}
\begin{definition}  \label{dmut}
(i) We define $\mu_t$ as the Borel probability measure on ${\cal H}_{\alpha,+}$
 which gives the distribution of the random solution $Y(t)$,
 \beqn\nonumber \mu_t(B) =
\mu_0(U_+(-t)B),\,\,\,\, B\in {\cal B}({\cal H}_{\alpha,+}),\,\,\,t\in \R\,.
 \eeqn
(ii) The correlation functions of the  measure $\mu_t$ are  defined by
\be\label{qd}
Q_t^{ij}(z,z')= E \Big(Y^i(z,t)\otimes
 Y^j(z',t)\Big),\,\,\,i,j= 0,1,\,\,\,\,z,z'\in\Z^d_+.
\ee
Here $Y^i(z,t)$ are the components of the random solution
$Y(t)=(Y^0(\cdot,t),Y^1(\cdot,t))$ to the problem (\ref{CP1}).
\end{definition}

The main result of the paper is the following theorem.
\medskip\\
{\bf Theorem A}
{\it  Let $d,n\ge 1$, $\alpha<-d/2$, and assume that the conditions 
(\ref{condE0}), {\bf E1}--{\bf E6} and {\bf S1}--{\bf S4} hold. Then\\
(i) the convergence in (\ref{1.8i}) holds. \smallskip\\
(ii) The limit measure $ \mu_\infty$ is a Gaussian
measure on ${\cal H}_{\alpha,+}$.\smallskip\\
(iii) The correlation matrices of the measures $\mu_t$
converge to a limit, for $i,j=0,1$,
\be\label{corf}
Q^{ij}_t(z,z')=\int\big( Y^i(z)\otimes Y^j(z')\big)
\,\mu_t(dY)\to Q^{ij}_\infty(z,z'),\,\,\,\,t\to\infty,\quad
z,z'\in\Z^d_+.
\ee
The correlation matrix $Q^{ij}_\infty(z,z')=(Q^{ij}_\infty(z,z'))_{i,j=0}^1$
of the limit measure $\mu_{\infty}$ has a form
\be\label{1.13}
Q_\infty(z,z')=q_\infty(z-z')-q_\infty(z-\tilde z')-q_\infty(\tilde z-z')+
q_\infty(\tilde z-\tilde z'),\quad z,z'\in\Z^d_+.
\ee
Here $q_\infty(z)=  q^+_{\infty}(z)+ q^-_{\infty}(z)$,
where in the Fourier transform we have
\beqn
 \hat q^+_{\infty}(\theta)&=&\frac{1}{4} \sum\limits_{\sigma=1}^s
\Pi_\sigma(\theta)\left(\hat {\bf q}_0(\theta) 
+C(\theta)\hat {\bf q}_0(\theta)C(\theta)^*\right)\Pi_\sigma(\theta),
\label{1.14}\\
\hat q^-_{\infty}(\theta)&=&\frac{i}{4}\sum\limits_{\sigma=1}^s
{\rm sign} \left(\partial_{\theta_1}\omega_\sigma(\theta)\right)
 \Pi_\sigma(\theta) 
 \left(C(\theta)\hat {\bf q}_0(\theta) -
\hat {\bf q}_0(\theta)C(\theta)^*\right)\Pi_\sigma(\theta),
\,\,\,\theta\in\T^d\setminus {\cal C}_*,\,\label{1.15}
\eeqn 
$\Pi_\sigma(\theta)$ is the spectral projection from Lemma \ref{lc*} (iv) and  
\be\label{C(theta)}
C(\theta)=\left(\ba{cc}
0&\Omega(\theta)^{-1}\\
-\Omega(\theta)&0 \ea\right)\,,\quad 
C(\theta)^*=\left(\ba{cc} 0&-\Omega(\theta)\\
\Omega(\theta)^{-1}&0 \ea\right).
 \ee
(iv) The measure $\mu_\infty$ is time stationary, i.e.,
$[U_+(t)]^*\mu_\infty=\mu_\infty$, $t\in\R$.\\
(v) The group $U_+(t)$ is mixing with respect to   
 the measure $\mu_\infty$, i.e.,   
$$   
\lim_{t\to\infty}   
\int f(U_+(t)Y)g(Y)\,\mu_{\infty}(dY)=\int f(Y)\,\mu_{\infty}(dY)
\int g(Y)\,\mu_{\infty}(dY)   
$$   
for any $f,g\in L^2({\cal H}_{\alpha,+},\mu_\infty)$. }
\medskip

The assertions {\it (i)}, {\it(ii)} of Theorem~A can be deduced
  from Propositions  \ref{l2.1} and \ref{l2.2} below.
 \begin{pro}\label{l2.1}
  The family of  measures $\{\mu_t,\,t\in \R\}$
 is weakly compact on the space ${\cal H}_{\alpha,+}$ for any  $\alpha<-d/2$,
and the bounds 
$\sup\limits_{t\ge 0} E \Vert U_+(t)Y_0\Vert^2_{\alpha,+} <\infty$ hold.
 \end{pro}

 Set ${\cal S}=[S(\Z^d_+)\otimes \R^n]^2$, where
$S(\Z^d_+)$ stands for the space of rapidly decreasing real sequences.
Denote $\langle Y,\Psi\rangle_+ =\langle Y^0,\Psi^0\rangle_+ +\langle
Y^1,\Psi^1\rangle_+$ for  $Y=(Y^0,Y^1)\in {\cal H}_{\alpha,+}$ and 
$\Psi=(\Psi^0,\Psi^1)\in  {\cal S}$,
where
$$
\langle Y^i,\Psi^i \rangle_+=\sum\limits_{z\in\Z^d_+}Y^i(z)\cdot\Psi^i(z),
\quad i=0,1.
$$
 \begin{pro}\label{l2.2}
  For every $\Psi\in {\cal S}$, the characteristic functionals 
converge to a Gaussian one, 
\be\label{2.6i}
 \hat\mu_t(\Psi): = \int e^{i\langle Y,\Psi\rangle_+}\mu_t(dY)
\rightarrow \exp\left\{-\frac{1}{2}{\cal Q}_\infty (\Psi ,\Psi)\right\},\,\,\,
t\to\infty,
 \ee
where ${\cal Q}_\infty$ is the quadratic form defined as
$$
{\cal Q}_\infty (\Psi ,\Psi)=\sum\limits_{i,j=0}^1
\sum\limits_{z,z'\in\Z^d_+}
 \Bigl(Q_\infty^{ij}(z, z'),\Psi^i(z)\otimes \Psi^j(z') \Bigr).
$$
 \end{pro}
Proposition  \ref{l2.1} ensures the existence of the limit measures
of the family $\{\mu_t,\,t\in \R\}$, while Proposition \ref{l2.2}
 provides  the uniqueness. They are proved in Sections 6 and 7, 
respectively. The assertion {\it (iii)} of Theorem~A is proved in Section 5,
item {\it(iv)} follows from (\ref{1.8i}) and item {\it(v)}
can be proved using a method of \cite{D2}. 
\begin{remarks}\label{remi-iii} 
{\rm   (i)
 The {\it uniform Rosenblatt mixing condition} \cite{Ros} also suffices, 
together with a higher power $>2$  in the bound (\ref{med}): 
there exists $\delta >0$ such that
\be\label{med'}
E \Big( \vert Y^0_0(z)\vert^{2+\delta}+\vert Y^1_0(z)\vert^{2+\delta}
\Big) \le C <\infty,\quad z\in\Z^d_+.
\ee
Then  (\ref{1.12}) requires a modification:
$$
\ds\int_0^{+\infty}\ds r^{d-1}\alpha^{p}(r)dr <\infty,\quad\mbox{with }\,
p=\min(\delta/(2+\delta),  1/2).
$$
Here $\alpha(r)$ is the Rosenblatt mixing coefficient  defined
as in  (\ref{ilc}) but without $\mu_0(B)$ in the denominator:
$$
\alpha(r)=\sup\{\alpha_Y({\cal A},{\cal B}):
\,{\cal A},{\cal B}\subset \Z^d_+,\,\,\dist({\cal A},\,{\cal B})\geq r\},
$$
where
$$
\alpha_Y({\cal A},{\cal B})=\sup\{|\mu_0(A\cap B)-\mu_0(A)\mu_0(B)|:\,
A\in\sigma({\cal A}),\,B\in\sigma({\cal B})\}.
$$
Under these modifications, the statements of Theorem A and their
proofs remain essentially unchanged.

(ii) The  uniform Rosenblatt mixing condition also could be weakened.
Let $K(z,s)=\prod\limits_{i=1}^d[z_i-s,z_i+s]$, where $s>0$, 
$z\in\Z^d$, stand for the cube in $\Z^d$,
$\bar K_r=\Z^d\setminus K(z,s+r)$. Let us define the mixing coefficient 
$\alpha_l(r)$ by the rule
$$
\alpha_l(r)=\sup\left\{
\alpha_Y\left(K(z,s)\cap \Z^d_+,\bar K_r\cap \Z^d_+\right):
\,z\in\Z^d_+,\,0\le s\le l\right\}.
$$
To prove Theorem A it suffices to assume, together with (\ref{med'}), that
$$
\alpha_l(r)\le \frac{C~l^\kappa}{(1+r)^{\kappa'}},
$$
with some constants $C,\kappa, \kappa'>0$.
See \cite{Bu, Do} for a more detailed discussion about the different
mixing conditions.

 (iii) The condition {\bf E5} could be considerably weakened.  
Namely, it suffices to assume the following condition:\\
{\bf E5'} If for some $\sigma\not=\sigma'$ one has
$\omega_\sigma(\theta)\pm\omega_{\sigma'}(\theta)\equiv \const_\pm$
with $\const_\pm\not=0$, then
\be\label{psisi'}
\left\{ \ba{rr}
p_{\sigma\sigma'}^{11}(\theta)\mp
\omega_{\sigma}(\theta)\omega_{\sigma'}(\theta)
p_{\sigma\sigma'}^{00}(\theta)\equiv 0,\\
\omega_\sigma(\theta)p_{\sigma\sigma'}^{01}(\theta)\pm
\omega_{\sigma'}(\theta)p_{\sigma\sigma'}^{10}(\theta)\equiv 0.
\ea\right.
\ee
Here $p_{\sigma\sigma'}^{ij}(\theta)$ stand for the matrices
$$
p_{\sigma\sigma'}^{ij}(\theta):=
\Pi_\sigma(\theta)\hat {\bf q}_0^{ij}(\theta)\Pi_{\sigma'}(\theta),\,\,\,\,
\theta\in \T^d,\,\,\,\, \sigma,\sigma'=1,\dots,s,\,\,\,\, i,j=0,1,
$$
 $\hat {\bf q}_0^{ij}(\theta)$
are Fourier transforms of the correlation functions ${\bf q}^{ij}_0(z)$.

Note that the condition {\bf E5'} is fulfilled, for instance, if
${\bf q}_0(z)$ is a covariance matrix of a Gibbs measure $\nu_0$ on
${\cal H}_\alpha$, $\alpha<-d/2$. 
Formally the Gibbs measure $\nu_0$ is 
$$
\nu_{0}(dv_0, dv_1)= \frac{1}{Z_\beta}
\ds e^{-\ds  \frac{\beta}{2}\ds
\sum_{z\in\Z^d} (|v_1(z)|^2 +{\cal V}v_0(z)\cdot v_0(z))}\prod_{z}
dv_0(z)dv_1(z),
$$ 
where $Z_\beta$ is the normalization factor, $\beta= T^{-1}$,
$T>0$ is an absolute temperature. 
Let us introduce the Gibbs measure $\nu_{0}$ 
as the Gaussian measure with the correlation matrices 
defined by their Fourier transform as
$$
\hat {\bf q}_0^{00}(\theta)= T\hat V^{-1}(\theta),~~~
\hat {\bf q}_0^{11}(\theta)= T 
\left(\delta_{kl}\right)_{k,l=1}^n,~~~
\hat {\bf q}_0^{01}(\theta)= \hat {\bf q}_0^{10}(\theta)= 0.
$$
Then $p_{\sigma\sigma'}^{ij}=0$ for 
$\sigma\not=\sigma'$ and (\ref{psisi'}) holds.}   
\end{remarks}   

\setcounter{equation}{0}
\section{Examples}
Let us give the examples of the equations (\ref{1+}) and measures $\mu_0$ 
which satisfy all conditions {\bf E1}--{\bf E6}  and {\bf S1}--{\bf S4}, 
respectively.

\subsection{Harmonic crystals}

The conditions {\bf E1}--{\bf E6} in particular are fulfilled 
in the case of {\it the nearest neighbor crystal}, i.e.,
when the interaction matrix $V(z)=(V_{kl}(z))_{k,l=1}^n$ has a form:
\be\label{V}
\ba{ccl}
V_{kl}(z)&=&0\mbox{ for }k\not=l,\\
V_{kk}(z)&=&\left\{\ba{ll}-\gamma_k&\mbox{for }\, |z|=1,\\
2\gamma_k+m_k^2& \mbox{for }\, z=0,\\
0&\mbox{for }\, |z|\ge2,\ea\right.  \quad k=1,\dots,n,
\ea
\ee
with $\gamma_k>0$, $m_k\ge0$.
Then the equation (\ref{1+}) becomes
$$
\ddot u_k(z,t)=(\gamma_k\Delta_L-m_k^2)u_k(z,t),\quad k=1,\dots,n.
$$
Here $\Delta_L$ stands for the discrete
Laplace operator on the lattice $\Z^d$,
$$
\Delta_L u(z):= \sum\limits_{e,|e|=1}(u(z+e)-u(z)).
$$
Then the eigenvalues of  $\hat V(\theta)$ are
 $$
 \tilde{\omega}_k(\theta)=
\sqrt{\, 2 \gamma_1(1-\cos\theta_1)+...+2 \gamma_d
(1-\cos\theta_d)+m_k^2}\,,\quad k=1,\dots,n.
 $$
These eigenvalues have to be labelled as follows
\beqn 
\tilde\omega_1(\theta)\equiv\dots\equiv\tilde\omega_{r_1}(\theta),\,\,\,   
\tilde\omega_{r_1+1}(\theta)\equiv\dots\equiv\tilde\omega_{r_2}(\theta),
\,\,\, \dots,\,\,\,\,
\tilde\omega_{r_{s-1}+1}(\theta)\equiv\dots\equiv \tilde\omega_{r_s}(\theta) 
\nonumber\\ 
\omega_1(\theta)\equiv \tilde\omega_{r_1}(\theta)<
\omega_2(\theta)\equiv \tilde\omega_{r_2}(\theta)<\dots<
\omega_s(\theta)\equiv \tilde\omega_{r_s}(\theta).
\nonumber
\eeqn 
Clearly conditions {\bf {E1}}--{\bf {E5}} hold with ${\cal C}_*=\emptyset$.
In the case all $m_k>0$ the set ${\cal C}_0$ is empty and
condition {\bf E6} holds automatically. Otherwise, if $m_k=0$ for
some $k$,  ${\cal C}_0=\{0\}$. Then {\bf E6} is equivalent to the 
condition $\omega_\sigma^{-2}(\theta)\in L^1(\T^d)$, which holds if $d\ge 3$. 
Therefore, the conditions {\bf E1}--{\bf E6} hold provided 
either (i) $d\ge 3$, or (ii) $d=1,2$ and all $m_k >0$. 

In the case (\ref{V}) formulas (\ref{1.14}) and (\ref{1.15}) 
can be rewritten as follows. Denote
$$
\chi_{kl}(\sigma)= \left\{ 
\ba{rl} 
1 &{\rm if} \,\,\,\, k,l\in[r_{\sigma-1}+1, r_{\sigma}]\\ 
 0& {\rm otherwise}  
\ea \right. \,\,\, \sigma=1,...,s, 
$$
Then
$$   
 \hat q^{ij}_{\infty\,kl}=\frac14\sum\limits_{\sigma=1}^s
\chi_{kl}(\sigma)M_{kl}^{ij},\quad i,j=0,1,\quad k,l=1,\dots,n,
$$
where
\beqn   
M_{kl}^{11}&=& \omega_\sigma^2 M_{kl}^{00}= 
\left[ \omega_\sigma^2 \hat {\bf q}_0^{00}+ \hat {\bf q}_0^{11} 
 -i\sign(\sin\theta_1)\omega_\sigma(\hat {\bf q}^{01}_0-\hat {\bf q}^{10}_0)
\right]_{kl},\nonumber \\
M_{kl}^{01}&=&-M_{kl}^{10}=\Big[\hat {\bf q}^{01}_0-\hat {\bf q}^{10}_0
+i\, \frac{\sign(\sin\theta_1)}{\omega_\sigma(\theta)}
\left(\omega_\sigma^2 \hat {\bf q}_0^{00}+ \hat {\bf q}_0^{11} \right)
\Big]_{kl}.\nonumber
 \eeqn 

\subsection{Gaussian measures}
We consider $n=1$ and construct
Gaussian initial measures $\mu_0$ satisfying {\bf S1}--{\bf S4}.
Let us define  $\nu_0$ in ${\cal H}_\alpha$ by the  correlation
functions ${\bf q}_0^{ij}(z-z')$ which are zero for
 $i\not= j$, while for $i=0,1$,
\be\label{S04}
\hat {\bf q}_0^{ii}(\theta):=F_{z\to\theta}
[ {\bf q}_0^{ii}(z)]\in L^1(\T^d),\,\,\,\,
\hat {\bf q}_0^{ii}(\theta) \ge 0.
\ee
Then by the Minlos theorem there exists a unique  Borel Gaussian measure
$\nu_0$ on  ${\cal H}_\alpha$, $\alpha<-d/2$,
with the correlation functions ${\bf q}^{ij}_0(z-z')$, because 
$$
\int\Vert X\Vert^2_\alpha\nu_0(dX)
=\sum\limits_{z\in\Z^d}(1+|z|^2)^\alpha
(\tr {\bf q}^{00}_0(0)\!+\!\tr {\bf q}_0^{11}(0))
=C(\alpha,d)\int\limits_{\T^d}
\tr( \hat {\bf q}^{00}_0(\theta)\!+\!\hat {\bf q}_0^{11}(\theta))\,d\theta
<\infty.
$$
The measure $\nu_0$ satisfies {\bf S1} and {\bf S3}.
Let us take a function $\zeta\in C(\Z)$ such that
$$
\zeta(s)= \left\{ \ba{ll}
1,~~\mbox{for }~ s>\,a,\\
0,~~\mbox{for }~ s\le0\ea\right.
\quad\mbox{with an }\, a>0.
$$
Let us introduce $X(z)$ as a random function in probability space
$({\cal H}_\alpha,\nu_0)$.
Define a Borel probability measure  $\mu_0$ on ${\cal H}_{\alpha,+}$
as a distribution of the random function
$Y_0(z)= \zeta(z_1)X(z)$, $z\in\Z^d_+$.
Then correlation functions of $\mu_0$ are
$$
Q_0^{ij}(z,z')=
{\bf q}_0^{ij}(z-z')\zeta(z_1)\zeta(z'_1),~~i,j= 0,1,
$$
where $z= (z_1,\dots,z_d)$, $z'= (z'_1,\dots,z'_d)\in \Z^d_+$,
and ${\bf q}_0^{ij}$ are the correlation functions  of the measure $\nu_0$. 
The measure $\mu_0$ satisfies {\bf S1}--{\bf S3}.
Further, let us provide, in addition to (\ref{S04}), that
\be\label{S5}
{\bf q}_0^{ii}(z)=0,\,\,\,|z|\geq r_0.
\ee
Then the mixing condition {\bf S4} follows with 
 $\varphi(r)=0$, $r\geq r_0$.
For instance, (\ref{S04}) and (\ref{S5}) hold if we set 
${\bf q}_0^{ii}(z)= f(z_1)f(z_2)\cdot\dots\cdot f(z_d)$,
 where $f(z)=N_0-|z|$ for $|z|\le N_0$ and $f(z)=0$ for
$|z|> N_0$ with $N_0:=[r_0/\sqrt d]$ (the integer part).
Then 
$ \hat f(\theta)=(1-\cos N_0\theta)/(1-\cos\theta)$,
$\theta\in \T^1$, and (\ref{S04}) holds.

\setcounter{equation}{0}
\section{Convergence of correlation functions}
\subsection{Bounds for initial covariance}
\begin{definition}
By $l^p\equiv l^p(\Z^d)\otimes \R^n$ 
$(l^p_+\equiv l^p(\Z^d_+)\otimes \R^n)$, $p\ge 1$, $n\ge 1$,
 we denote the space of sequences
$f(z)=(f_1(z),\dots,f_n(z))$ endowed with norm
$\Vert f\Vert_{l^p}=\Big(\sum\limits_{z\in\Z^d}|f(z)|^p\Big)^{1/p}$
($\Vert f\Vert_{l^p_+}:=\Big(\sum\limits_{z\in\Z^d_+}|f(z)|^p\Big)^{1/p}$,
resp.).
\end{definition}
The next Proposition reflects the mixing property 
of initial correlation functions. 
\begin{pro} \label{l4.1}
(i) Let  conditions {\bf S1}--{\bf S4} hold. Then
for $i,j=0,1$, the following bounds hold
\beqn
\sum\limits_{z'\in\Z^d_+} |Q^{ij}_0(z,z')|
&\le& C<\infty\,\,\,\mbox{ for all }\,z\in\Z^d_+,
\label{pr1}\\
\sum\limits_{z\in\Z^d_+} |Q^{ij}_0(z,z')|
&\le& C<\infty\,\,\,\mbox{ for all }\,z'\in\Z^d_+.
\label{pr2}
\eeqn
Here the constant $C$ does not depend on $z,z'\in \Z^d_+$.\\
(ii)  $\hat {\bf q}^{ij}_0\in  C(\T^d)$, $i,j=0,1$.
\end{pro}
{\bf Proof}.
(i) By \cite[Lemma 17.2.3]{IL}, 
conditions {\bf S1}, {\bf S3} and {\bf S4} imply 
\be\label{4.9'}
|Q^{ij}_0(z,z')| \le C e_0\,\varphi^{1/2}(|z-z'|),~~ z,z'\in\Z^d_+.
\ee
Hence, condition (\ref{1.12}) implies (\ref{pr1}),
\be\label{qp}
\sum\limits_{z\in\Z^d_+}|Q^{ij}_0(z,z')| \le C e_0
 \sum\limits_{z\in\Z^d} \varphi^{1/2}(|z|) <\infty.
\ee
(ii) The bound (\ref{4.9'}) and condition {\bf S2} imply
the following bound:
\be\label{4.9}
|{\bf q}^{ij}_0(z)|\le C e_0\,\varphi^{1/2}(|z|),~~ z\in\Z^d.
\ee
Hence, from (\ref{1.12}) it follows that ${\bf q}^{ij}_0(z)\in l^1$,
what  implies $\hat {\bf q}^{ij}_0\in C(\T^d)$.\bo
\begin{cor}\label{c4.10}
Proposition \ref{l4.1} (i) implies, by the Shur lemma,
 that for any $\Phi,\Psi\in l^2_+$ the following bound holds:
$$
|\langle Q_0(z,z'),\Phi(z)\otimes\Psi(z')\rangle_+|\le
C\Vert\Phi\Vert_{l^2_+} \Vert\Psi\Vert_{l^2_+}.
$$
\end{cor}

\subsection{Proof of the convergence (3.2)}
From condition (\ref{condE0}), formulas (\ref{Grcs}) and (\ref{hA})
it follows that ${\cal G}_t(z)={\cal G}_t(\tilde z)$
with $\tilde z=(-z_1,z_2,\dots,z_d)$. Then, by the explicit representation
(\ref{sol}), the covariance $Q_t(z,z')$ can be decomposed into a sum
of four terms:
$$
Q_t(z,z')=\sum\limits_{y,y'\in\Z^d_+}{\cal G}_{t,+}(z,y)Q_0(y,y')
{\cal G}^T_{t,+}(z',y')
=R_t(z,z')-R_t(z,\tilde z')-R_t(\tilde z,z')+R_t(\tilde z,\tilde z'),
$$
where 
$$
R_t(z,z'):=\sum\limits_{y,y'\in\Z^d_+}
{\cal G}_t(z-y)Q_0(y,y'){\cal G}^T_t(z'-y').
$$
Therefore, (\ref{corf}) follows from the following convergence
\be\label{5.6}
R_t(z,z')\to q_\infty(z-z')\quad \mbox{as }\,t\to\infty,\quad
z,z'\in\Z^d.
\ee
To prove (\ref{5.6}) let us define 
$$
Q_*(z,z')=\left\{
\ba{cl}
Q_0(z,z')& \mbox{for }\,z,z'\in\Z^d_+,\\
0& \mbox{otherwise}.
\ea\right. 
$$ 
First we split the function $Q_*(z,z')$ into the following three matrices
\beqn
Q^+(z,z')&:=&\frac12{\bf q}_0(z-z'),\label{d1'}\\
Q^-(z,z')&:=&\frac12{\bf q}_0(z-z')\sign (z'_1),\label{d1''}\\
Q^r(z,z')&:=&Q_*(z,z')-Q^+(z,z')-Q^-(z,z').\label{d1'''}
\eeqn
Next introduce the  matrices
\be\label{Qta}
R^a_{t}(z,z')= \sum\limits_{y,y'\in\Z^d}
\Big({\cal G}_t(z-y) Q^a(y,y')
{\cal G}_t^T(z'-y')\Big),\,\,\,\,z,z'\in \Z^d,\,\,\,\,t>0,
\ee
for each $a=\{+,-,r\}$, and split
$R_t(z,z')$ into three terms:
$R_t(z,z')= R^+_{t}(z,z')+R^-_{t}(z,z')+R^r_{t}(z,z')$.
Then the convergence (\ref{5.6}) follows from the following lemma. 
\begin{lemma}\label{Qt1}
(i) $\lim\limits_{t\to\infty} R_t^+(z,z')= q^+_\infty(z-z')$,
$z,z'\in\Z^d$, 
with the matrix $q^+_{\infty}$ defined in (\ref{1.14}),\\
(ii) $\lim\limits_{t\to\infty} R_t^-(z,z')= q^-_\infty(z-z')$,
$z,z'\in\Z^d$,  
with the matrix $q^-_{\infty}$  defined in (\ref{1.15}).\\
(iii) $\lim\limits_{t\to\infty} R_t^r(z,z')=0$, $z,z'\in\Z^d$. 
\end{lemma}
This lemma can be proved using the technique of \cite[Proposition 7.1]{DKM}.
To justify the main idea of the proof we sketch the proof 
of Lemma \ref{Qt1} (i) in Appendix.

\setcounter{equation}{0}
\section{Compactness of measures family}

Proposition \ref{l2.1} follows from the bound (\ref{20.1})
by the Prokhorov compactness theorem \cite[Lemma II.3.1]{VF}
by a method used in \cite[Theorem XII.5.2]{VF},
since the embedding  ${\cal H}_{\alpha,+}\subset {\cal H}_{\beta,+}$
is compact if $\alpha>\beta$.
\begin{lemma}\label{lcom}
Let conditions {\bf S1}, {\bf S3}, {\bf S4} hold and $\alpha<-d/2$.
Then  the following bounds hold
 \beqn \label{20.1}
\sup\limits_{t\ge 0}
E\Vert U_+(t)Y_0\Vert^2_{\alpha,_+}<\infty.     
\eeqn
\end{lemma}
{\bf Proof}. Definition \ref{d1.1} implies
$$
E \Vert  Y(\cdot,t)\Vert^2_{\alpha,_+}=
\!\sum\limits_{z\in \Z^d_+} (1+|z|^2)^\alpha
\Big({\rm tr}\,Q_t^{00}(z,z)+{\rm tr}\,Q_t^{11}(z,z)\Big)<\infty.
$$
Since  $\alpha<-d/2$, it remains to prove that
$$
\sup\limits_{t\in\R} \sup\limits_{z,z'\in \Z^d_+}
\Vert Q_t(z,z')\Vert\le C<\infty.
$$
The representation (\ref{sol}) gives
\beqn
Q^{ij}_t(z,z')&=&E\Big(Y^i(z,t)\otimes Y^j(z',t)\Big)
= \sum\limits_{y,y'\in \Z^d_+} \sum\limits_{k,l=0,1}
{\cal G}^{ik}_{t,+}(z,y)Q^{kl}_0(y,y'){\cal G}^{jl}_{t,+}(z',y')\nonumber\\
&=& \langle Q_0(y,y'), \Phi^i_{z}(y,t)\otimes
\Phi^j_{z'}(y',t)\rangle_+,\nonumber
\eeqn
where
\beqn
\Phi^i_{z}(y,t)&:=&\Big(
{\cal G}^{i0}_{t,+}(z,y),{\cal G}^{i1}_{t,+}(z,y)\Big)\nonumber\\
&=&({\cal G}_t^{i0}(z-y)-{\cal G}_t^{i0}(z-\tilde y),
{\cal G}_t^{i1}(z-y)-{\cal G}_t^{i1}(z-\tilde y)),
\,\,\,\,\,i=0,1.\nonumber
\eeqn
Note that the Parseval identity, formula (\ref{hatcalG})
and condition {\bf E6} imply
$$
\Vert\Phi^i_{z}(\cdot,t)\Vert^2_{l^2}= (2\pi)^{-d}
\int\limits_{\T^d} |\hat\Phi^i_{z}(\theta,t)|^2\,d\theta
\le C\int\limits_{\T^d}
\Big( |\hat{\cal G}^{i0}_t(\theta)|^2
+|\hat{\cal G}^{i1}_t(\theta)|^2\Big)
\,d\theta \le C_0<\infty.
$$
Then Corollary \ref{c4.10} gives
$$
|Q^{ij}_t(z,z')|=
|\langle Q_0(y,y'), \Phi^i_{z}(y,t)\otimes
\Phi^j_{z'}(y',t)\rangle_+|
\le C\Vert\Phi^i_{z}(\cdot,t)\Vert_{l^2_+}\,
\Vert\Phi^j_{z'}(\cdot,t)\Vert_{l^2_+}\le C_1<\infty,
$$
where the constant $ C_1$  does  not depend on
$z,z'\in\Z^d_+$, $t\in\R$. \bo

\setcounter{equation}{0}
\section{Convergence of characteristic functionals}
We derive (\ref{2.6i}) by using the explicit representation (\ref{sol}) 
of the solution $Y(t)$, the Bernstein `room - corridor' technique
and a method of \cite{DKKS}--\cite{DKM}. The method gives a representation 
of $\langle Y(t),\Psi\rangle_+$ as a sum of weakly dependent random variables 
(see formula (\ref{razli}) below). 
Then (\ref{2.6i})  follows from the central 
limit theorem under a Lindeberg-type condition. 
The similar technique of the proof is applied in \cite[Sections 9, 10]{DKM}.
Then we remark only the main steps of the proof.

\subsection{Asymptotics of $U'_+(t)\Psi$}

At first, let us evalute of scalar product $\langle Y(t),\Psi\rangle_+$.
Let us introduce a function $\Psi_*(z)$ as
$$
\Psi_*(z)=\left\{\ba{cl}
\Psi(z),&\mbox{if }\,z_1>0,\\
0,&\mbox{if }\,z_1=0,\\
-\Psi(\tilde z),&\mbox{if }\,z_1<0.
\ea\right.
$$
Therefore 
\beqn\label{YP}
\langle Y(z,t),\Psi(z)\rangle_+=\langle Y(z,t),\Psi_*(z)\rangle_+=
\langle Y_0(z'),\Phi(z',t)\rangle_+,
\eeqn
where 
\beqn\label{7.2}
\Phi(z',t)&:=&U'_+(t)\Psi_*(z')
=\sum\limits_{z\in\Z^d_+} {\cal G}_{t,+}^{T}(z,z')\Psi_*(z)
=\sum\limits_{z\in\Z^d} {\cal G}_{t}^{T}(z-z')\Psi_*(z)\nonumber\\
&=&(2\pi)^{-d}\int\limits_{\T^d} e^{-iz'\cdot\theta}\hat{\cal G}^*_t
(\theta)\hat\Psi_*(\theta)\,d\theta.
\eeqn
\begin{definition}\label{dC}
(i) The critical set ${\cal C}:={\cal C}_0\cup{\cal C}_*
\cup\Big(\cup_1^s{\cal C}_\sigma\Big)$ (see {\bf E4}).\\
(ii) ${\cal S}^0:=\{\Psi\in{\cal S}=[S(\Z^d)\otimes\R^n]^2:
\hat\Psi(\theta)=0\,\, \mbox{\rm in a neighborhood of}\,\,{\cal C}\}$.
\end{definition}

Note that {\rm mes}\,${\cal C}=0$ (see \cite[lemma 7.3]{DKM})
and  it suffices to prove (\ref{2.6i}) for $\Psi_*\in {\cal S}^0$ only. 
For the function $\Phi(z,t)$ the following lemma holds.
\begin{lemma}\label{l5.3} (cf Lemma 9.1 from \cite{DKM})
Let conditions {\bf E1}--{\bf E4} and {\bf E6} hold. Then
for any fixed $\Psi_*\in {\cal S}^0$,  the following bounds hold:\\
 (i) $\sup_{z\in\Z^d}|\Phi(z,t)| \le  C~t^{-d/2}$.\\
(ii) For  any $p>0$ there exist $C_p,\gamma>0$ such that 
\be\label{conp}
|\Phi(z,t)|\le C_p(1+|z|+|t|)^{-p},\quad |z|\ge\gamma t.
\ee
\end{lemma}
This lemma follows from (\ref{7.2}), (\ref{hatcalG}), Definition \ref{dC} 
and the standard stationary phase method.

\subsection{Bernstein's argument}

Let us  introduce a `room - corridor'  partition of the
half-ball $\{z\in\Z^d_+:~|z|\le \gamma t\}$,  with $\gamma$ from (\ref{conp}).
For $t>0$ we choose $\Delta_t$ and $\rho_t\in\N$.
Let us choose  $0<\delta<1$ and
\be\label{rN}
\rho_t\sim t^{1-\delta},
~~~\Delta_t\sim\frac t{\log t},~~~~\,\,\,t\to\infty.
\ee
Let us set $h_t=\Delta_t+\rho_t$ and   
$$   
a^j=jh_t,\,\,\,b^j=a^j+\Delta_t,\,\,\,   
j=0,1,2,\dots,\,N_t=[(\gamma t)/h_t].   
$$   
We call the slabs $R_t^j=\{z\in\Z^d_+:|z|\le N_t h_t,\,a^j\le z_1< b^j\}$ 
the `rooms',   
$C_t^j=\{z\in\Z^d_+: |z|\le N_t h_t,\, b^j\le z_1<a^{j+1}\}$ the `corridors'   
and $L_t=\{z\in\Z^d_+: |z|> N_t h_t\}$ the 'tail'.   
Here  $z=(z_1,\dots,z_d)$, $\Delta_t$ is the width of a room, and   
$\rho_t$  of a corridor. Let us denote  by   
 $\chi_t^j$ the indicator of the room $R_t^j$, 
 $\xi_t^j$ that of the corridor $C_t^j$, and  
$\eta_t$ that of the tail $L_t$. Then  
$$
{\sum}_t [\chi_t^j(z)+\xi_t^j(z)]+ \eta_t(z)=1,\,\,\,z\in\Z^d_+,  
$$   
where the sum ${\sum}_t$ stands for $\sum\limits_{j=0}^{N_t-1}$.   
Hence, we get the following  Bernstein's type representation:   
\be\label{res}   
\langle Y_0,\Phi(\cdot,t)\rangle_+ = {\sum}_t   
\left[\langle Y_0,\chi_t^j\Phi(\cdot,t)\rangle_+ +   
\langle Y_0,\xi_t^j\Phi(\cdot,t)\rangle_+ \right]+   
\langle Y_0,\eta_t\Phi(\cdot,t)\rangle_+.   
\ee   
Let us define  the random variables $r_{t}^j$, $c_{t}^j$, $l_{t}$ by   
$$
r_{t}^j= \langle Y_0,\chi_t^j\Phi(\cdot,t)\rangle_+,~~   
c_{t}^j= \langle Y_0,\xi_t^j\Phi(\cdot,t)\rangle_+,   
\,\,\,l_{t}= \langle Y_0,\eta_t\Phi(\cdot,t)\rangle_+.   
$$   
Therefore, from (\ref{YP}) and (\ref{res}) it follows that 
\be\label{razli}   
\langle Y(t),\Psi\rangle_+=\langle Y_0,\Phi(\cdot,t)\rangle_+ =   
{\sum}_t (r_{t}^j+c_{t}^j)+l_{t}.   
\ee   
\begin{lemma}  \label{l5.1} 
    Let {\bf S1}--{\bf S4} hold and $\Psi_*\in{\cal  S}^0$.
The following bounds hold for $t>1$:
\beqn
E|r^j_{t}|^2&\le&  C(\Psi)~\Delta_t/ t,\,\,\,\forall j,\nonumber\\
E|c^j_{t}|^2&\le& C(\Psi)~\rho_t/ t,\,\,\,\forall j,\nonumber\\
E|l_{t}|^2&\le& C_p(\Psi)~t^{-p},\,\,\,\,\forall p>0.
\nonumber
\eeqn
\end{lemma}
The proof is based on Lemma \ref{l5.3} and Proposition \ref{l4.1} (i) 
(see \cite[Lemma 9.2]{DKM}).
\medskip

Further, to prove (\ref{2.6i}) we use a version of the central limit theorem
developed by Ibragimov and Linnik.
If  ${\cal Q}_{\infty}(\Psi,\Psi)=0$, the convergence (\ref{2.6i}) follows
from (\ref{corf}).
Thus, we may assume that for a given $\Psi_*\in{\cal S}^0$,
\be\label{5.*}
{\cal Q}_{\infty}(\Psi,\Psi)\not=0.
\ee
At first, we obtain 
$$
| E\exp\{i \langle Y_0,\Phi(\cdot,t)\rangle_+\} - \hat\mu_{\infty}(\Psi)|
=\left|E\exp\left\{i{\sum}_t r_t^j\right\}
-\exp\left\{-\frac12 {\sum}_t E|r_t^j|^2\right\}\right|+o(1),\,\,\,t\to\infty.
$$
This fact follows from Lemma \ref{l5.1}, convergence (\ref{corf}),
condition {\bf S4} and (\ref{rN}) (cf \cite[p.1073-1075]{DKM}).
 
Secondly, by the mixing condition {\bf S4}, we derive that
$$
\left|E\exp\left\{i{\sum}_t r_t^j\right\}-\prod\limits_{0}^{N_t-1}
E\exp\left\{i r_t^j\right\}\right|
\le C N_t\varphi(\rho_t)\to 0,\quad t\to\infty.
$$
Hence, it remains to check that
$$
\left|\prod\limits_{0}^{N_t-1} E\exp\left\{ir_t^j\right\}
-\exp\left\{-\frac12{\sum}_{t} E|r_t^j|^2\right\}\right| \to 0,~~t\to\infty.
$$
According to the standard statement of the central limit theorem
(see, e.g. \cite[Theorem 4.7]{P}), 
it suffices to verify the  Lindeberg condition:
$\forall\delta>0$,
$$
\frac{1}{\sigma_t}{\sum}_t  E_{\delta\sqrt{\sigma_t}}
|r_t^j|^2 \to 0,~~t\to\infty.
$$
Here $\sigma_t\equiv {\sum}_t E |r^j_t|^2$,
and $E_\ve f\equiv E (X_\ve f)$,
where $X_\ve$ is the indicator of the event $|f|>\ve^2$.
Note that (\ref{corf}) and (\ref{5.*}) imply  that
$\sigma_t \to{\cal Q}_{\infty}(\Psi, \Psi)\not= 0$, $t\to\infty$.
Hence it remains to verify that
$$
{\sum}_t E_{\ve} |r_t^j|^2 \to 0,~~t\to\infty,
 ~~ \mbox{ for any }\, \ve>0.
$$
This condition is checked using the technique from \cite[section 10]{DKM}.
\bo

\setcounter{equation}{0}
\section{Appendix. Outline of the proof of Lemma 5.4 (i)}
Obviously, the assertion of Lemma \ref{Qt1} (i) 
is equivalent to the next proposition.
\begin{pro} 
Let conditions {\bf E1}--{\bf E6} and {\bf S1}--{\bf S4} hold.
Then for any $\Psi\in{\cal S}$,
\be\label{8.1}
\lim\limits_{t\to\infty}\langle R_t^+(z,z'),\Psi(z)\otimes\Psi(z')\rangle
= \langle q^+_\infty(z-z'),\Psi(z)\otimes\Psi(z')\rangle.
\ee
\end{pro}
{\bf Proof}. It suffices to prove (\ref{8.1}) for $\Psi\in{\cal S}^0$ only.
It can be proved similarly as in \cite[Lemma 7.6]{DKM}.

At first, let us apply the Fourier transform to the matrix
$R_{t}^+(z,z')$ defined by (\ref{Qta}):
$\hat R^{+}_{t}(\theta,\theta'):=
F\!\!\!_{\scriptsize {\ba{ll}z\to\theta\\ z'\to \theta'
\ea}}\!\! R^{+}_{t}(z,z')=
\hat {\cal G}_t(\theta)\hat Q^{+}(\theta,\theta')\hat {\cal G}_t^T(\theta')$,
where
$\hat Q^{+}(\theta,\theta'):=F\!\!\!_{\scriptsize {\ba{ll}
z\to\theta\\ z'\to \theta' \ea}}\!\! Q^+(z,z')$.
From  (\ref{d1'}) it follows that
$\hat Q^+(\theta,\theta')=
\delta(\theta+\theta')~(2\pi)^{d}~\hat{\bf q}_0(\theta)/2$.
Hence, 
$$
\hat R^+_{t}(\theta,\theta')=(2\pi)^d\frac12\delta(\theta+\theta')
\hat {\cal G}_t(\theta)\hat{\bf q}_0(\theta) \hat{\cal G}_t^T(-\theta).
$$
Secondly, $\hat{\cal G}_t( \theta)$ has a form
\be\label{hatcalG}
\hat{\cal G}_t( \theta)=
\left( \begin{array}{cc}
 \cos\Omega t &~ \sin \Omega t~\Omega^{-1}  \\
 -\sin\Omega t~\Omega
&  \cos\Omega t\end{array}\right),
\ee
where $\Omega=\Omega(\theta)$ is the Hermitian matrix defined by (\ref{Omega}).
Let $C(\theta)$ be defined by (\ref{C(theta)})
and  $I$ be the identity matrix. Then
\be\label{Gtdec}
\hat{\cal G}_t( \theta)=\cos\Omega t\, I+\sin\Omega t\, C(\theta).
\ee
Moreover, by condition {\bf E2}, 
$\hat {\cal G}_t^T(-\theta)=\hat {\cal G}_t^*(\theta)=
\cos\Omega t\, I+\sin\Omega t\, C(\theta)^*$. Therefore,
\beqn\label{Qt,1}
\langle R_t^+(z,z'),\Psi(z)\otimes\Psi(z')\rangle
&=&(2\pi)^{-2d}\langle \hat R_t^+(\theta,\theta'),\hat \Psi(\theta)
\otimes\hat\Psi(\theta')\rangle
\nonumber\\
&=&\frac1{2(2\pi)^{d}}\langle\hat {\cal G}_t(\theta)
\hat {\bf q}_0(\theta)\hat {\cal G}_t^*(\theta),
\hat \Psi(\theta)\otimes\ov{\hat \Psi}(\theta)\rangle.
\eeqn
Further, let us choose certain smooth branches of the functions 
$\Pi_\sigma(\theta)$ and $\omega_\sigma(\theta)$ to apply the stationary phase 
arguments which require a smoothness in $\theta$.
We  choose  a finite partition of unity 
\be\label{part} 
\sum_{m=1}^M g_m(\theta)=1,\,\,\,\,\theta\in \supp\hat\Psi, 
\ee 
where $g_m$ are nonnegative functions from  $C_0^\infty(\T^d)$ and    
vanish  in a neighborhood of the set ${\cal C}$
  defined in Definition \ref{dC}, (i).
Further, using (\ref{part}) we rewrite the  RHS of (\ref{Qt,1}).
Applying formula (\ref{Gtdec}) for $\hat {\cal G}_t(\theta)$,
  one obtains
$$
\langle R_t^+(z,z'),\Psi(z)\otimes\Psi(z')\rangle
=\frac1{2(2\pi)^{d}}\sum_{m}\sum\limits_{\sigma,\sigma'=1}^s
\int\limits_{\T^d}  g_m(\theta) 
\Big(\Pi_\sigma(\theta) R_{t,\sigma\sigma'}(\theta) \Pi_{\sigma'}(\theta),
\hat \Psi(\theta)\otimes\ov{\hat \Psi}(\theta)\Big)\,d\theta,
$$
where $R_{t,\sigma\sigma'}(\theta)$ stands for the $2n\times 2n$ matrix,
\beqn\label{7.9}
R_{t,\sigma\sigma'}(\theta) &=&\frac{1}{2}\sum\limits_{\pm}\Big\{
\cos\big(\omega_\sigma(\theta)\pm\omega_{\sigma'}(\theta)\big)t
\Big[\hat {\bf q}_0(\theta)\mp C(\theta)\hat {\bf q}_0(\theta)
 C(\theta)^*\Big]
\nonumber\\
&&+\sin\big(\omega_\sigma(\theta)\pm\omega_{\sigma'}(\theta)\big)t
~\Big[ C(\theta)\hat {\bf q}_0(\theta)\pm\hat {\bf q}_0(\theta)
C(\theta)^*\Big]\Big\}.
\eeqn 
If $\sigma=\sigma'$, then
\beqn\label{8.7}
R_{t,\sigma\sigma}(\theta) &=&\frac{1}{2}\Big[\hat {\bf q}_0(\theta)+
 C(\theta)\hat {\bf q}_0(\theta) C(\theta)^*\Big]
+\frac12\cos\big(2\omega_\sigma(\theta)t\big)\Big[\hat {\bf q}_0(\theta)-
 C(\theta)\hat {\bf q}_0(\theta) C(\theta)^*\Big]\nonumber\\
&&+\frac12\sin\big(2\omega_\sigma(\theta)t\big)
\Big[C(\theta)\hat {\bf q}_0(\theta)+ \hat {\bf q}_0(\theta)
 C(\theta)^*\Big].
\eeqn
By Lemma \ref{lc*} and the compactness arguments, 
we choose the eigenvalues $\omega_\sigma(\theta)$ 
and the matrix $\Pi_\sigma(\theta)$ as real-analytic functions inside 
the $\supp g_m$ for every $m$: we do not mark the 
functions by the index $m$ to not overburden the notations.
Let us analyze the Fourier integrals with   $g_m$.

At first, note that the identities  
$\omega_\sigma(\theta)+\omega_{\sigma'}(\theta)\equiv\const_+$ 
or $\omega_\sigma(\theta)-\omega_{\sigma'}(\theta)\equiv\const_-$
 with the $\const_\pm\ne 0$ are impossible by   condition {\bf E5}.
Furthermore, the oscillatory integrals 
with $\omega_\sigma(\theta)\pm \omega_{\sigma'}(\theta)\not\equiv \const$
vanish as $t\to\infty$. Hence, 
 only the integrals with  
$\omega_\sigma(\theta)-\omega_{\sigma'}(\theta)\equiv 0$
contribute to the limit,
since  $\omega_\sigma(\theta)+\omega_{\sigma'}(\theta)\equiv 0$ would imply  
$\omega_\sigma(\theta)\equiv\omega_{\sigma'}(\theta)\equiv 0$ which 
is impossible by  {\bf E4}.
By  formulas (\ref{7.9}) and (\ref{8.7}), one obtains
\beqn
&&\langle R_t^+(z,z'),\Psi(z)\otimes\Psi(z')\rangle\nonumber\\
&=& (2\pi)^{-d}\sum\limits_m\sum\limits_{\sigma=1}^s\frac14
\int\limits_{\T^d} g_m(\theta)\Bigl(\Pi_\sigma(\theta)
\Big[\hat {\bf q}_0(\theta)+
 C(\theta)\hat {\bf q}_0(\theta) C(\theta)^*\Big]
\Pi_\sigma(\theta)+\dots,
\hat \Psi(\theta)\otimes\ov{\hat \Psi}(\theta)\Big)\,d\theta
\nonumber\\
&=&(2\pi)^{-d}\int\limits_{\T^d}
\Bigl(\hat q^+_{\infty}(\theta),
\hat \Psi(\theta)\otimes\ov{\hat \Psi}(\theta)\Big)\,d\theta 
+\dots, \nonumber
\eeqn
where  $"\dots"$ stands for the oscillatory  integrals
which contain  $\cos(\omega_\sigma(\theta)\pm\omega_{\sigma'}(\theta))t$ 
and $\sin(\omega_\sigma(\theta)\pm\omega_{\sigma'}(\theta))t$ 
with $\omega_\sigma(\theta)\pm\omega_{\sigma'}(\theta)\not\equiv$const. 
The oscillatory integrals converge to zero  by the Lebesgue-Riemann theorem  
since all the integrands in `$...$'  are summable, and  
$\nabla(\omega_\sigma(\theta)\pm\omega_{\sigma'}(\theta))=0$ only on the set 
of the Lebesgue measure zero. 
The summability follows from Proposition \ref{l4.1} (ii)
and  {\bf E6} (if ${\cal C}_0\not=\emptyset$)  
 since the matrices $\Pi_\sigma(\theta)$ are bounded. 
The zero measure follows similarly to Lemma \ref{lc*} (i) 
since $\omega_\sigma(\theta)\pm\omega_{\sigma'}(\theta)\not\equiv\const$.  
Lemma \ref{Qt1} (i) is proved. 
\bo   
\medskip 
\begin{center}
{\bf Acknowledgment}
\end{center}

Author would like to thank  Prof. H. Spohn for the helpful discussions.


\end{document}